\documentclass[prb,twocolumn,superscriptaddress]{revtex4-1}

\usepackage{graphicx}

\usepackage[english]{babel}
\usepackage[utf8]{inputenc} 

\usepackage{amsthm,dsfont,float,amsfonts,amsmath}

\begin{document}

\title{Thermoelectric effects in graphene with local spin-orbit interaction}
\author{M. I. Alomar}
\affiliation{Institut de F\'{\i}sica Interdisciplin\`aria i Sistemes Complexos IFISC (UIB-CSIC), E-07122 Palma de Mallorca, Spain}
\affiliation{Departament de Física, Universitat de les Illes Balears, E-07122 Palma de Mallorca, Spain}
\author{David S\'anchez} 
\affiliation{Institut de F\'{\i}sica Interdisciplin\`aria i Sistemes Complexos IFISC (UIB-CSIC), E-07122 Palma de Mallorca, Spain}
\affiliation{Departament de Física, Universitat de les Illes Balears, E-07122 Palma de Mallorca, Spain}

\begin{abstract}
We investigate the transport properties of a graphene layer in the presence of Rashba spin-orbit
interaction. Quite generally, spin-orbit interactions induce spin splittings
and modifications of the graphene bandstructure. We calculate within the scattering approach
the linear electric and thermoelectric responses of a clean sample when the Rashba coupling
is localized around a finite region. We find that the thermoelectric conductance, unlike its electric counterpart,
is quite sensitive to external modulations of the Fermi energy. Therefore, our results suggest
that thermocurrent measurements may serve as a useful tool to detect nonhomogeneous spin-orbit
interactions present in a graphene-based device. Furthermore, we find that the junction thermopower
is largely dominated by an intrinsic term independently of the spin-orbit potential scattering.
We discuss the possibility of cancelling the intrinsic thermopower by resolving the Seebeck coefficient
in the subband space. This causes unbalanced populations of electronic modes 
which can be tuned with external gate voltages or applied temperature biases.
\end{abstract}

\maketitle

\section{Introduction}\label{sec:int}

Graphene is a single layer of carbon atoms arranged on a two-dimensional honeycomb lattice.\cite{Geim1,Geim2}
The study of its electronic properties has recently attracted great interest \cite{Castro,GN,AKG,Beenakker} in part
due to peculiar features of its energy bandstructure. Within a tight-binding model, graphene's conduction and valence bands touch each other at six different points, the $K$-points, which reduce to two, $K$ and $K^{\prime}$, because the rest are equivalent by symmetry. Near these points and at low energies, electrons behave as massless fermions travelling at fixed velocity $V_F \sim 10^6$~m/s, independently of their energy. Then, the energy spectrum consists of two cones that come into contact
at their vertices and the low-energy excitations can be conveniently described by an effective Dirac-Weyl equation
where the speed of light is replaced with $V_F$.\cite{Castro}

Recent works suggest large spin-orbit strengths in graphene layers under the influence
of metallic substrates. \cite{var08,ded08,ras09,ert09,abd10,li11}
This finding is interesting in view of recent studies that relate spin-orbit coupling
of the Rashba type \cite{ras1,ras2} to topological insulating behavior.\cite{KaneMele,KaneMele2}
Importantly, the Rashba coupling strength can be externally tuned by modifying the electric field applied to a nearby gate. \cite{Nitta} This type of interaction leads to band splittings and enriched spintronic effects. \cite{mor99,mir01} In semiconductor quantum wires with parabolic confinement,
the presence of localized Rashba interaction has been predicted
to yield Fano antiresonances, \cite{san06,jeo06,zha06,per07,she08,san08}
to help the detection of entangled electrons, \cite{egu02,pab06,maz13}
and to assist electron-spin resonance manipulation. \cite{fro09,sad13,hac13}
The effect of nonhomogeneous Rashba couplings has also been considered
in the transport characteristics of two-dimensional systems. \cite{mat02,kho04,pal04,ora10,gel11}
A natural question is thus to ask to what extent these results are modified
in graphene monolayers. In contrast to semiconductor heterojunctions, in graphene electrons are massless and the spin-orbit interaction depends on a pseudospin degree, not a momentum.

\begin{figure}[t]
\centering
\includegraphics[width=0.45\textwidth]{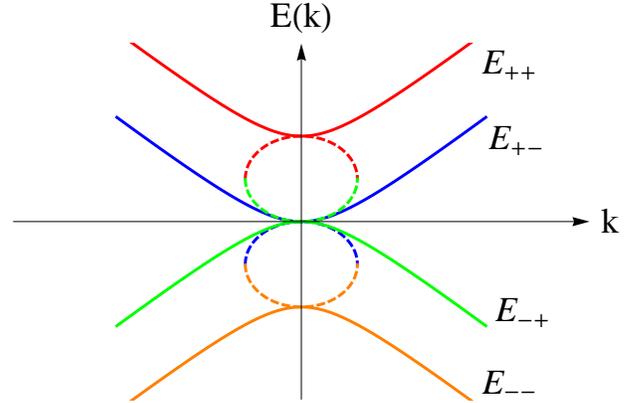}
\caption{(Color online) Sketch of the energy bandstructure of a graphene layer with spin-orbit interaction of the Rashba type. 
Solid lines indicate propagant states while dashed lines depict the energy associated to evanescent states.
Due to the spin-orbit potential, four bands (labeled as $++$, $+-$, $-+$, and~$--$) are obtained.\label{fig:ISO}}
\end{figure}

In this paper, we investigate the influence of local Rashba spin-orbit interaction
on the electric and thermoelectric properties of graphene. We shall focus on the linear regime
of transport. Previous studies have considered Fano lineshapes in graphene junctions, \cite{Yamakage,rat11}
spin densities in nanoribbons \cite{Stauber} and superlattices,\cite{Shakouri}
spin dependent transmissions \cite{Bercioux} and Klein (chiral) tunneling. \cite{liu12}
Here, we are mainly concerned with the voltages generated in response
to a temperature difference (the Seebeck effect). \cite{Ashcroft}
Interestingly, recent results indicate enhanced thermopower in graphene 
monoloyers, \cite{dra07,zue09,wei09} which paves the way
for promising applications to achieve efficient heat-to-energy converters. \cite{she12}
We here discuss the possibility of manipulating the thermopower with
a local spin-orbit interaction. In fact, we find that a spin-orbit
graphene monolayer is more sensitive to temperature biases than to voltage
differences. Furthermore, since the Rashba coupling splits the graphene electronic bandstructure
(see Fig.~\ref{fig:ISO}),
the transmission thus depends on the subband index.
In analogy with spin caloritronic devices, \cite{bau10} where a thermal gradient
induces a spin-polarized voltage bias, \cite{Uchida,jaw10} we propose to use the Seebeck
effect to generate a difference between occupations with different subband indices.

The paper is organized as follows. In Sec. \ref{sec:model}, we describe our model Hamiltonian and investigate
the effects of a uniform spin-orbit coupling in both the energy spectrum and the electronic states
of a flat graphene sheet. Section~\ref{sec:scat}
describes the system under consideration: a junction with a Rashba interaction potential localized
around a central region where the spin-orbit strength is nonzero and constant. Using matching methods
for wave functions with four components, we find the transmission probabilities for all incident electronic modes.
Importantly, the transmission function shows for a given subband index a critical angle beyond which 
electrons cannot be transmitted across the junction. The electric conductance and the subband
polarization are discussed in Sec.~\ref{sec:con}. We find that the polarization rapidly changes in the energy scale
of the Rashba strength for sufficiently wide spin-orbit regions. Section~\ref{sec:the} contains the central results
of our work. We calculate the thermocurrent in response to a small temperature shift and obtain strong
modulations when the Fermi energy is tuned even to values much larger than the spin-orbit strength.
Surprisingly, the Seebeck coefficient is a smooth function of energy, an effect which we attribute
to a background intrinsic thermopower which is dominant for a wide range of Fermi energies.
We then determine the subband thermovoltage generated in response to a temperature bias
and recover the strong variation with energy, yielding positive or negative population imbalances
depending on the value of the externally tuned Fermi energy.
Finally, our conclusions are summarized in Sec. \ref{sec:conclu}.

\begin{figure}[t]
\centering
\includegraphics[width=0.47\textwidth]{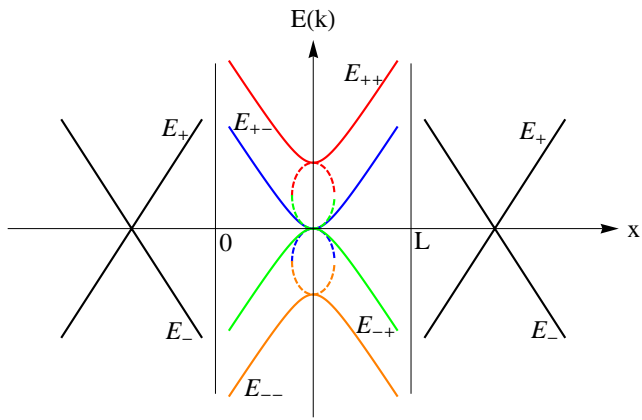}
\caption{(Color online) Pictorial representation of a graphene layer with a central region of length $L$ where
spin-orbit interaction is active. We take $x$ as the propagation direction.
We show the energy spectra both inside and outside the central region.
\label{fig:noISOsiISOnoISO2d}}
\end{figure}

\section{Theoretical model}\label{sec:model}

We consider a graphene layer in the $xy$ plane with spatially varying spin-orbit interaction
along the $x$ direction. Within the continuum limit, the total Hamiltonian reads
\begin{equation}\label{eq:h}
{\cal H}=-i \hbar V_F (\sigma_x  {\partial}/{\partial x}+ \sigma_y  {\partial}/{\partial y})\otimes s_o 
+\lambda ( \sigma_x \otimes s_y - \sigma_y \otimes s_x)\,.
\end{equation}

The first term in the right-hand side represents the effective Hamiltonian for electrons in a clean
graphene sheet. This model is valid when inter-valley scattering can be safely neglected.
The electron spin and pseudospin (sublattice) degrees of freedom are taken into account with
the Pauli $s$ and $\sigma$ matrices, respectively. The second term describes
the Rashba spin-orbit coupling with $\lambda$ the interaction strength.\cite{KaneMele}
We take $\lambda$ as a slowly varying function in a length scale larger than
the graphene lattice constant. Thus, the continuum model remains valid.
Furthermore, in Eq.~\eqref{eq:h} we have neglected the intrinsic contribution to the spin-orbit
interaction since this term is much smaller than the Rashba coupling and cannot be
externally tuned.\cite{hue06,min06}

Let $k$ ($q$) be the wavevector component along the $x$ ($y$) direction.
Then, the eigenenergies of ${\cal H}$ are given by
\begin{equation}
E_{l,n}=l \sqrt{\lambda ^2+\hbar ^2 V_F^2 (k^2+q^2)}+n \lambda \,,\label{eq:e2}
\end{equation}
where $l=\pm$ labels states with positive or negative energies and
$n=\pm$ is the subband index. The energy spectrum is plotted
with solid lines in Fig.~\ref{fig:ISO} for a finite value of the spin-orbit
strength $\lambda$. The energy bands split with a splitting given
by $2\lambda$ for both the positive and negative branches of the spectrum.

The eigenstates of ${\cal H}$ are
\begin{equation}
\psi_{l,n}^m(x)= \frac{ e^{i m k x}  e^{i qy}/2}{\sqrt{\hbar ^2 V_F^2 (k^2+q^2)+E_{l,n}^2} }
\left(\begin{array}{c}
-in \hbar V_F(m k-i q)\\
E_{l,n} \\
 -i n E_{l,n}\\
\hbar V_F( m k+i q)
\end{array}\right)\,,  \label{eq:f2}
\end{equation}
where we explicitly indicate the propagation direction with the aid of the index $m=\pm$,
which determines the sign of the momentum along $x$. 
Since the scattering potential is invariant in the $y$ direction we take $q$ as a real quantity. However,
the $k$ momentum can be real or purely imaginary depending on whether one deals with
traveling or evanescent waves. A systematic method of finding evanescent states in
quantum wires with Rashba interaction is presented in Ref.~\onlinecite{ser07}. Here, we notice that the energy
of evanescent waves emerges from the subband spectra and coalesces for $E=\pm\lambda$
(see the dashed lines in Fig.~\ref{fig:ISO})

In the absence of spin-orbit interaction, we recover the well known dispersion relation for bare graphene,
\begin{equation}
E_l=l \hbar V_F \sqrt{k^2+q^2} \,,\label{eq:e13}
\end{equation}
with eigenstates
\begin{equation}
\psi_{l,n}^m(x,y)= \frac{ e^{i m k x}  e^{i q y}}{2}
\left(\begin{array}{c}
-inm e^{-i m \phi}\\
l \\
 -i n l\\
 m e^{i m \phi}
\end{array}\right) \,. \label{eq:f13}
\end{equation}
Here, $\phi$ is the wavevector angle defined as $\phi=\tan^{-1}q/k$. We represent Eq.~\eqref{eq:e13}
for $q=0$ in the left and right sides of Fig.~\ref{fig:noISOsiISOnoISO2d}. The spectrum $E(k)$ is linear
with a constant slope. In contrast, in the presence of Rashba interaction the energy bands become parabolic
for energies small compared to the spin-orbit strength (central area in Fig.~\ref{fig:noISOsiISOnoISO2d}). 

\section{Local Rashba interaction}\label{sec:scat}

We investigate the scattering problem sketched in Fig.~\ref{fig:noISOsiISOnoISO2d} with three distinct regions.
While the side regions (left and right) are bare graphene, the central region of length $L$
is subjected to spin-orbit interaction of the Rashba type. 
Since the problem is invariant in the direction perpendicular to $x$,
the $y$-component of the momentum does not change and we can write it in terms of the wavevector angle,
\begin{equation}\label{eq:q}
q=\frac{E}{\hbar V_F} \sin\phi\,.
\end{equation}

We consider electrons with fixed energy $E>0$. From Eqs.~(\ref{eq:e2}) and (\ref{eq:q}),
we obtain the wavevector component parallel to the transport direction,
\begin{equation}\label{eq:k13}
k=E\sqrt{(1-\sin^2\phi)}/\hbar V_F\,,
\end{equation}
valid for $x<0$ and $x>L$. For $0<x<L$, $k$ can be determined from Eqs.~(\ref{eq:e2}) and (\ref{eq:e13}):
\begin{equation}\label{eq:k2}
k^n=\sqrt{E(E-2 n\lambda-E \sin^2\phi)}/\hbar V_F\,.
\end{equation}

In the central region, we have two possible values for $k^n$, one for subband with $n=+$ and one for subband with $n=-$,
although a more careful analysis is needed
in terms of the subband index. First, we notice that, in general, for any energy the momentum is always real if
$E-2n \lambda-E \sin^2\phi>0$, i.e.,
\begin{equation}\label{eq:cond}
\sin\phi<\sqrt{\frac{E-2 n \lambda}{E}}\,.
\end{equation}

Now, for $E>2\lambda$ and $n=-$, the Eq.~(\ref{eq:cond}) is always satisfied since $\sin\phi$ is bounded between $0$ and $1$.
In contrast, for $n=+$ we have a critical angle at which the momentum becomes pure imaginary. For angles higher than the critical angle we have an evanescent wave.  Then, for $0<E<2\lambda$ and $n=-$, the Eq.~(\ref{eq:cond}) also holds
as before, but for $n=+$ the momentum becomes pure imaginary since Eq.~(\ref{eq:cond}) is never satisfied and the wave is evanescent for any value of the angle $\phi$. Similar critical angles have been  invoked to discuss total internal reflection effects in semiconductor interfaces with spin orbit interaction.\cite{kho04} 

\begin{figure}[t]
\centering
\includegraphics[width=0.47\textwidth]{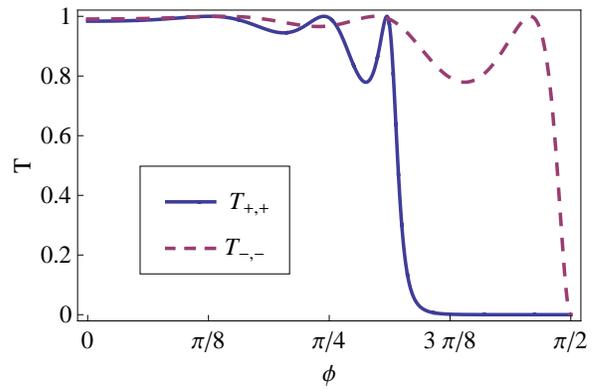}
\caption{(Color online) Transmission probability as a function of  the incident angle $\phi$ for $E>2\lambda$. The  solid curve represents the trasmision from $n=+$ to $n=+$ and the dashed curve the trasmision from $n=-$ to $n=-$. Parameters: $\lambda=10$ meV, $E=80$ meV and $L=100$ nm. \label{fig:TmaEma2L}}
\end{figure}

We are now in a position to solve the scattering problem in Fig.~\ref{fig:noISOsiISOnoISO2d}.
We focus on the case $E>0$ since our system exhibits particle-hole symmetry, even in the presence
of Rashba coupling. Therefore, we take $l=+1$.
We consider a most simple inhomogeneity, namely,
$\lambda=0$ for $x<0$ and $x>L$, and $\lambda$ nonzero and uniform for $0<x<L$. This is not contradictory with the assumption that $\lambda$ is a slowly varying function in an atomic level, because the scale over which this change takes place is much bigger than the graphene lattice constant.
The matching method allows us to calculate all reflection and transmission amplitudes
for a given electron, which we take as impinging from the left. In the following,
we express the wave function at each region as $\psi^{m}_{+,n}(x,y)=\psi_{n}^m e^{imkx}e^{iqy}$ [cf. Eq.~\eqref{eq:f2}].
We first specify left ($\ell$) wave function for $x<0$:
\begin{eqnarray}
\psi_{\ell,n}(x,y)&=&\psi_{n}^+e^{i k x}  e^{i qy}+ r_{n,-}\psi_{-}^-e^{-i k x}  e^{i q y} \nonumber \\
&& +r_{n,+}\psi_{+}^-e^{-i k x}  e^{i q y}\,,
\end{eqnarray}
where the incident subband $n$ can be taken as $+$ or $-$. The reflection amplitudes $r_{n,-}$ and $r_{n,+}$
describe back scattering into $-$ and $+$ modes, respectively.
Then, we have an incident wave with positive group velocity, $v=k>0$, and two reflected  waves with $v=-k<0$,
the latter belonging to the doubly degenerate $ E_+ $ branch in Fig.~\ref{fig:noISOsiISOnoISO2d}.

In the central ($c$) region we have four coexisting waves,
\begin{eqnarray}\label{eq:f2}
\psi_{c,n}(x,y)&=&a_{n,-}\psi_{-}^+e^{i k^- x}  e^{i qy}+b_{n,+}\psi_{+}^+e^{i k^+ x}  e^{i qy} \nonumber\\
&+&c_{n,-}\psi_{-}^-e^{-i k^- x}  e^{i qy}+d_{n,+}\psi_{+}^-e^{-i k^+ x}  e^{i qy}\,,
\end{eqnarray}
where the coefficients $a$, $b$, $c$, and $d$ are labeled with the incident subband $n$
and the wavevector index $\pm$ defined in Eq.~\eqref{eq:k2}. Note that the propagating
or evanescent character of the partial waves is determined by the real or imaginary value of $k^\pm$. Equation ~\eqref{eq:f2} is valid for $E>\lambda$, but for $0<E<\lambda$ we need to take into account the evanescent states taking $\psi_{+}^+$ and $\psi_{+}^-$ for $l=-1$.

Finally, in the right ($r$) region we only have transmitted waves with positive group velocity
and positive and negative $n$:
\begin{equation}
\psi_{r,n}(x,y)=t_{n,-}\psi_{-}^+e^{i k x}  e^{i qy}+t_{n,+}\psi_{+}^+e^{i k x}  e^{i qy}\,,
\end{equation}
where $t_{n,\pm}$ denotes the transmission amplitude from the $n$-th incident subband toward the $\pm$ mode.

At the boundaries $x=0$ and $x=L$ we impose continuity of the wave function,
\begin{eqnarray}
\psi_{\ell,n}(0,y)&=&\psi_{c,n}(0,y) \,, \\ 
\psi_{c,n}(L,y)&=&\psi_{r,n}(L,y)  \,,
\end{eqnarray}
from which the eight coefficients $r_{n,\pm}$, $a_{n,-}$, $b_{n,+}$, $c_{n,-}$, $d_{n,+}$ and $t_{n,\pm}$ are determined.

In elastic scattering, the probability current is conserved. Since our system shows scattering along $x$ only,
the current conservation condition leads to
\begin{equation}
1+R_{n,+}+R_{n,-}=T_{n,+}+T_{n,-}\,,
\end{equation}
where $ R_{n,\pm}=|r_{n,\pm} |^2$ ($ T_{n,\pm}=|t_{n,\pm} |^2$) is the reflection (transmission) probability.
Due to the spin-chiral nature of the carriers,\cite{Shakouri}
the off-diagonal probabilities $T_{+,-}$ and $T_{-,+}$, vanish altogether
and the spin-orbit interaction does not couple states with opposite subband indices.
Figure~\ref{fig:TmaEma2L} shows $ T_{+,+}$ and $T_{-,-}$ for $E>2 \lambda$ as a function of the incident angle.
At low angles ($\phi\simeq 0$) the transmission is close to unity. This is a manifestation of Klein tunneling
in graphene for incident wave vectors parallel to the transport direction.\cite{kas05}
When $\phi$ rotates from $0$, the transmission departs from $1$ due to scattering at the boundaries.
The situation is akin to a single-barrier potential\cite{kas05} but in our case the effect originates from
a purely spin-orbit field.

Interestingly, in Fig.~\ref{fig:TmaEma2L} we can see the emergence of a critical angle for $ T_{+,+}$ beyond which the transmission probability vanishes (solid line). It occurs because when we surpass the critical angle given by  Eq.~(\ref{eq:cond}), the wave into the central region becomes evanescent and the transmission drops. The transition is not abrupt since there are tunneling contributions
to $ T_{+,+}$ but this effect is very weak. Note that $ T_{-,-}$ (dashed line) does not show any critical angle, as predicted by
Eq.~(\ref{eq:cond}). Additionally, we also observe in Fig.~\ref{fig:TmaEma2L} transmission resonances which
we attribute to central waves interfering constructively for specific values of the incident angles. 

\begin{figure}[t]
\centering
\includegraphics[width=0.47\textwidth]{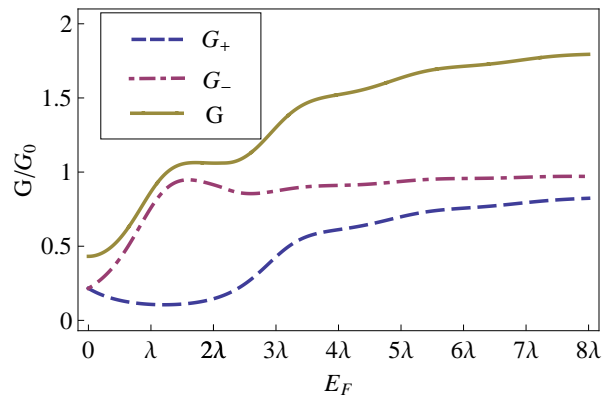}
\caption{(Color online) Conductance as a function of Fermi energy. Parameters: $\lambda=10$ meV and $L=100$ nm. \label{fig:G}}
\end{figure}

\section{Electric Conductance}\label{sec:con}
Within the scattering approach, the electric current carried by electrons in subband $n$ is obtained from
the transmission probabilities integrated over the injecting energies $E$ and the wave vector angle $\phi$,
\begin{eqnarray}\label{eq_In}
I_n&=&\frac{2 e W}{\pi h} \int_{0}^{\pi/2} \cos\phi\, d\phi \int_{-\infty}^{\infty} \mathcal{K}(E)\, T_{n,n}(E,\phi)  \nonumber \\ 
&\times &[ f_{L}(E) - f_{R}(E) ]\,dE\,,
\end{eqnarray}
where $W$ is the  sample width in the $y$ direction and $f_{\ell}(E)$ and $f_{r}(E)$ are Fermi-Dirac distribution functions
that describe the electronic population both in the left and right side, asymptotically far from the scattering (central) region. The $2$ factor is due to the valley degeneracy.
In Eq.~\eqref{eq_In}, $\mathcal{K}(E)=E/\hbar V_F$ is obtained from the graphene dispersion relation, Eq.~(\ref{eq:e13}). 
The total current is thus $I=\sum_n I_n$

To obtain the linear conductance $G=(dI/dV)_{V=0}$, a small voltage bias $V$ is applied across the junction.
We can shift the left Fermi-Dirac distribution $f_\ell=f(E-eV)$ fixing the right one $f_r=f(E)$,
where $f(E)=1/(1+e^{(E-E_F)/k_B T_0})$ is the equilibrium distribution function with
$E_F$ the Fermi energy and $T_0$ the background temperature.
After Taylor expanding Eq.~\eqref{eq_In} up to first order in $V$,
we find at zero temperature $G=\sum_nG_n$, where
\begin{equation}\label{eq:G}
G_n=G_0 \int^{\pi/2}_{0}  T_{n,n}(E_F,\phi) \cos\phi\,d\phi\,,
\end{equation}
$G_0=2 e^2 W {\cal K}_F/ \pi h=4 e^2 W E_F/h^2 V_F$ is
the maximum conductance of an ideal two-dimensional conductor since
${\rm Int}\,(W{\cal K}_F/\pi)$ is the number of open channels of a sample
with Fermi wave number $\mathcal{K}_F$.\cite{bee91}

Figure~\ref{fig:G} shows the conductance as a function of $E_F$.
We choose the Fermi energy as the changing parameter since it can
be easily tuned in an experimental setup.\cite{Geim1} The conductance $G_+$ (blue, dashed line) is small
for energies between $0$ and $2\lambda$. This is because in this energy range electrons from subband $+$
can be transmitted only via conventional tunnelling effect due to the center energy splitting.
The transmission probability is thus small. For $E_F$ higher than $2\lambda$, 
$G_+$ increases since travelling waves are  now permitted in the central region. However, the increase is slow due
to the persistence of the critical angle above which the transmission probability is zero. 
The conductance for the $-$ subband ($G_-$) is always close to the maximal value $G_0$ because for this mode there always
exists a travelling wave in the central region. In general, the total
conductance $G=G_++G_-$ (solid line) is a monotonically increasing function of the Fermi energy and tends to $2G_0$
at large energies where spin-orbit scattering is less efficient.
\begin{figure}[t]
\centering
\includegraphics[width=0.47\textwidth]{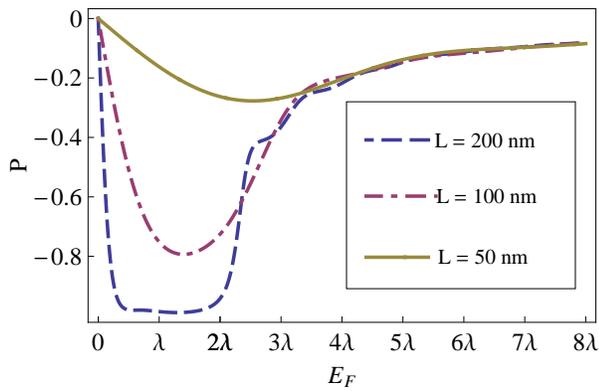}
\caption{(Color online) Subband polarization of the conductance as a function of Fermi energy for  $\lambda=10$ meV
and different values of the spin-orbit region length. \label{fig:P}}
\end{figure}

In Fig.~\ref{fig:P} we represent the subband polarization defined as
\begin{equation}\label{eq:P}
P=\frac{G_+-G_-}{G_++G_-}\,.
\end{equation}

We can see that for $\lambda < E_F< 2\lambda$ most of the electrons have negative polarization because for those energies the wave with positive $n$ becomes evanescent inside the central region and the transmission probability is very small.
This effect is more visible for wider regions of the spin-orbit stripe.
As we increase the Fermi energy, there are more electrons with positive polarization since for $E_F>2\lambda$
the states with $n=+$ are travelling waves and their transmission probability is larger.
Clearly, in the limit $E_F\gg \lambda$ electron scattering is insensitive
to the spin-orbit potential and the distinction between $+$ and $-$ subbands vanishes, yielding $P\to 0$.

\section{Thermoelectric conductance}\label{sec:the}

The current generated in the linear regime in response to a small temperature difference $\Delta T$ applied
across the junction can be obtained from Eq.~\eqref{eq_In} replacing the left Fermi-Dirac distribution with $f (E,T_0+\Delta T)$
and the right one with $f(E,T_0)$:
\begin{eqnarray}\label{eq_In2}
I_n&=&\frac{2 e W}{h \pi} \frac{\Delta T}{T_0} \int^{\pi/2}_{0} \cos\phi\, d\phi \int \frac{E}{\hbar V_F}(E-E_F) \nonumber \\
&\times& \left(-\frac{\partial f}{\partial E}\right) T_{n,n}(E,\phi) dE                                                                                              \,.
\end{eqnarray}

We are interested in the low temperature regime. Then, to leading order in a Sommerfeld expansion, the thermoelectric
conductance reads $\mathcal{L}=I/\Delta T=\sum_n \mathcal{L}_n$, where
\begin{eqnarray}\label{eq:L}
\mathcal{L}_n&=&\sum_{n}\mathcal{L}_0  \left[  \int^{\pi/2}_{0}\!\!\!\!\!\! T_{n,n} (E_F,\phi) \cos\phi \, d\phi \right. \nonumber\\
&+& \left. E_F \frac{\partial}{\partial E_F} \int^{\pi/2}_{0} \!\!\!\!\!\! T_{n,n}(E_F,\phi) \cos\phi \, d\phi \right]\,,
\end{eqnarray}
where $\mathcal{L}_0= k_B^2 e W T_0/3 \hbar^2 V_F$.

\begin{figure}
\centering
\includegraphics[width=0.47\textwidth]{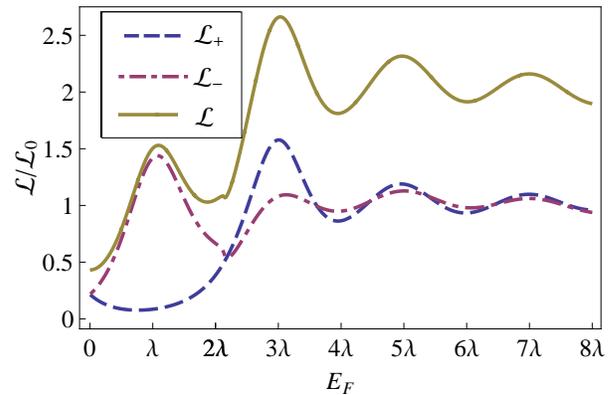}
\caption{(Color online) Thermoelectric conductance as a function of Fermi energy. Parameters: $\lambda=10$ meV
and $L=100$~nm. \label{fig:L}}
\end{figure}

In Fig.~\ref{fig:L}, we represent the thermoelectric conductance as a function of the Fermi energy. 
Surprisingly, we observe strongly modulated oscillations with a decreasing amplitude as we increase $E_F$.
This implies that the thermocurrent is more sensitive than the electric current to small variations of $E_F$.
Furthermore, we find that the position difference between consecutive peaks in $\mathcal{L}$
is approximately given by the spin-orbit strength $\lambda$. Therefore, thermoelectric measurements
can be rather useful in the detection of local spin-orbit fields in graphene single layers.

By virtue of the Seebeck effect, we expect that a thermovoltage will be generated
when the junction is in the presence of a temperature gradient under open circuit conditions.\cite{Ashcroft}
To keep our discussion general, we consider different electrochemical potentials $\mu_{\alpha n}=E_F+e V_n$ for each subband,
where $\alpha=\ell,r$. Both sides of the junction are maintained at different temperatures, $T_\alpha$,
independently of $n$. Then, the current flowing in the $n$ mode
in response to small shifts $ \mu_{\ell n}-\mu_{r n}$ and $T_\ell-T_r$ is
\begin{equation}
I_n=\frac{\mu_{\ell n}-\mu_{r n}}{e} G_n+(T_\ell-T_r) {\cal L}_n\,,
\end{equation}
where the transport coefficients $G_n$ and ${\cal L}_n$ are given by Eqs.~\eqref{eq:G} and~\eqref{eq:L}, respectively.

We define\cite{Rejec}
\begin{subequations}\label{eq_def}
\begin{eqnarray}
\Delta T&=&T_l-T_r\\
\mu_\alpha &=& \frac{1}{2}(\mu_{\alpha +}+\mu_{\alpha -})\\
eV&=&\mu_\ell -\mu_r\\
eV_s&=&(\mu_{\ell +}-\mu_{\ell -})-(\mu_{r +}-\mu_{r -})\,,
\end{eqnarray}
\end{subequations}
where $\Delta T$ is the temperature difference, $V$ the bias voltage and $V_s$ the subband voltage
that takes into account possible voltage differences in the same lead between different subbands.\cite{jon14}
Using Eq.~\eqref{eq_def} in Eqs.~\eqref{eq_In} and~\eqref{eq_In2}, we find the total current
\begin{equation}
I=(G_+\!+\!G_-)V\!+\!\frac{1}{2}(G_+\!-\!G_-)V_s\!+\!(\mathcal{L}_+\!+\!\mathcal{L}_-)\Delta T \,,\label{eq:IS}
\end{equation}
and the subband current $I_s=I_+-I_-$,
\begin{equation}
I_s=(G_+\!-\!G_-)V\!+\!\frac{1}{2}(G_+\!+\!G_-)V_s\!+\!(\mathcal{L}_+\!-\!\mathcal{L}_-)\Delta T \,.\label{eq:ISS}
\end{equation}

We note that $I_s$ is a polarization current in the subband space.
It then plays the role analogous to a spin or pseudospin current
since $n$ can take on two values only.

In Eqs.~\eqref{eq:IS} and~\eqref{eq:ISS}, the transport coefficients
are given by Eqs.~\eqref{eq:G} and~\eqref{eq:L}. Defining the integrated
transmission per subband as
\begin{equation}
\mathcal{T}_n(E_F)=\int^{\pi/2}_{0} T_{n,n} (E_F,\phi) \cos\phi  \,d\phi\,,
\end{equation}
Eqs.~\eqref{eq:G} and~\eqref{eq:L} can be recast in the form
\begin{subequations}\label{eq_gnln}
\begin{eqnarray}
G_n&=&\left(\frac{e}{\pi \hbar}\right)^2 \frac{W E_F  }{ V_F}  \mathcal{T}_n(E_F)\,, \\
\mathcal{L}_n&=&\frac{e k_B^2}{3 \hbar^2}\frac{W T_0}{ V_F}  \left[ \mathcal{T}_n(E_F) + E_F \frac{\partial \mathcal{T}_n}{\partial E_F} \right]\,,
\end{eqnarray}
\end{subequations}
where $\partial \mathcal{T}_n/\partial E_F$ is the energy derivative of $\mathcal{T}_n$
evaluated at $E_F$.

Interestingly, the low-temperature conductance is given by the integrated transmission, in agreement
with the Landauer picture of transport, but the thermoelectric conductance contains an additional term.
This can be seen more clearly in the calculation of the charge thermopower or Seebeck coefficient
$S=(V/\Delta T)_{I=0,V_s=0}$, which determines the voltage generated in the junction
in response to a temperature shift when the total current and the subband voltage are set to zero.
From Eqs.~(\ref{eq:IS}) and (\ref{eq:ISS}) we find
\begin{equation}\label{eq_sold}
S=-\left(\frac{{\cal L}_++{\cal L}_-}{G_++G_-}\right) \,.
\end{equation}

Inserting Eqs.~\eqref{eq_gnln} in Eq.~\eqref{eq_sold}, we obtain the low-temperature thermopower
\begin{equation}\label{eq_s}
S=-\frac{\pi^2 k_B}{3e}\frac{k_B T_0}{E_F}\left(1+E_F \frac{\sum_n\partial \mathcal{T}_n/\partial E_F}
{\sum_n \mathcal{T}_n} \right)\,.
\end{equation}

We notice two contributions in Eq.~\eqref{eq_s}. The second term in brackets can be
understood with the aid of the Mott formula $S\propto \partial \ln G/\partial E_F$, which is expected
to hold in generic conductors at low temperature. It is a single-particle result which is satisfied
in low dimensional systems such as quantum dots\cite{dot} and quantum point contacts,\cite{qpc} for which
a sizable thermopower is detected only if the transmission strongly depends on energy.
It is thus a pure transport contribution. However, Eq.~\eqref{eq_s} shows an additional term which is 
insensitive to transmission modulations. In fact, for a constant transmission
probability or when $\mathcal{T}_n$ shows a weak variation with energy on the scale of $E_F$,
Eq.~\eqref{eq_s} reduces to
\begin{equation}\label{eq_s2}
S\simeq -\frac{\pi^2 k_B}{3e}\frac{k_B T_0}{E_F}\,.
\end{equation}

This intrinsic contribution is independent of the sample details
and, more importantly, survives in the purely ballistic limit.
It simply states that in the highly degenerate limit ($E_F\gg T_0$, i.e., the range of validity
of the Sommerfeld approximation), the thermopower is given by the entropy per unit
charge ($k_B/e$) associated to the fraction of the electron density which is thermally
excited ($k_B T_0/E_F$). Therefore, Eq.~\eqref{eq_s2} is completely general and does not depend
on the nature of the scattering potential. For $E_F=1$~meV and $T_0=1$~K,
a typical value for the intrinsic thermopower
yields $S=20$~$\mu$V/K, a value detectable with present techniques.\cite{zue09}

\begin{figure}[t]
\centering
\includegraphics[width=0.47\textwidth]{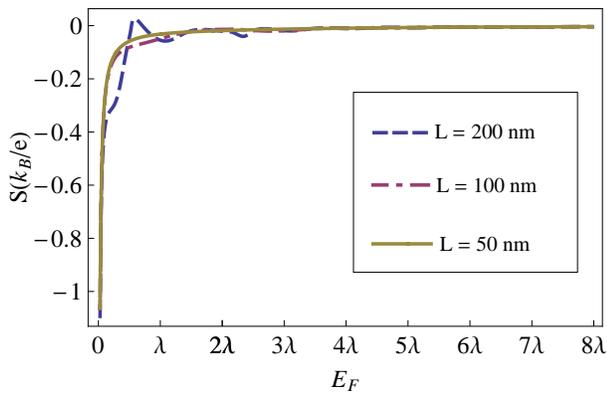}
\caption{(Color online) Seebeck coefficient as a function of
Fermi energy for different values of the spin-orbit region length.
Parameters: $\lambda=10$ meV and $T_0=1$ K.\label{fig:S}}
\end{figure}

We thus expect a competition between the intrinsic and the transport terms in the Seebeck coefficient.
We plot $S$ in Fig.~\ref{fig:S} as a function of $E_F$ for a nonzero value of the Rashba strength.
We observe that the junction thermopower is always negative, indicating that when the left side is hotter than
the right side, the system generates a negative bias to compensate the excess of thermally activated
electrons. Furthermore, $S$ is quite robust to variations of the spin-orbit region size $L$.
Additionally, the overall shape of $S$ is rather smooth unlike the strong variation of the thermoelectric
conductance as $E_F$ increases [cf. Fig.~\ref{fig:L}]. These facts can be explained
taking into account the intrinsic thermopower written in Eq.~\eqref{eq_s2}. At high energies the Rashba interaction
is not effective and the transmissions are weak functions of energy, as discussed in Sec.~IV.
Then, the transport contribution to the Seebeck coefficient, $S\propto \partial \ln G/\partial E_F$,
is negligible and $S$ tends to zero as $1/E_F$. At low energies the constant term exceeds the
transport contribution due to the $k_B T_0/E_F$ term.\cite{note} Therefore, the transport term
is relevant only at intermediate energies, as shown in Fig.~\ref{fig:S}.

Applied temperature gradients can also lead to spin accumulations in the attached leads,
as recently demonstrated in systems driven by spin Seebeck effects.\cite{Uchida,jaw10}
Then, it is natural to ask whether a local spin-orbit interaction in graphene leads
to different \textit{subband} populations. We address this question by calculating
from Eqs.~(\ref{eq:IS}) and~(\ref{eq:ISS})
the subband bias $V_s$ generated when $I=0$ and $I_s=0$ but $\Delta T\neq 0$.
The subband thermopower $S_s=V_s/\Delta T$ then follows, 
\begin{equation}\label{eq_Ss}
S_s=-\left( \frac{\mathcal{L}_+}{G_+}-\frac{\mathcal{L}_-}{G_-}\right)  \,.
\end{equation}

Notice that to obtain this result, we need to apply a bias voltage:
\begin{equation}
V=-\frac{1}{2}\left( \frac{\mathcal{L}_+}{G_+}+\frac{\mathcal{L}_-}{G_-}\right)\Delta T  \,.
\end{equation}

At low temperature, we can substitute Eq.~\eqref{eq_gnln} in Eq.~\eqref{eq_Ss}, yielding
\begin{equation}\label{eq_Ss2}
S_s=-\frac{\pi^2 k_B^2 T_0}{3e} \left(
\frac{\partial\mathcal{T}_+/\partial E_F}{\mathcal{T}_+}
-\frac{\partial\mathcal{T}_-/\partial E_F}{\mathcal{T}_-}
\right)\,.
\end{equation}

\begin{figure}[t]
\centering
\includegraphics[width=0.47\textwidth]{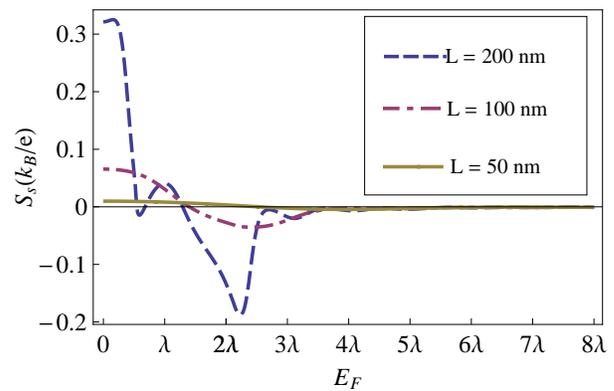}
\caption{(Color online) Subband-Seebeck coefficient as a function of
Fermi energy for different values of the spin-orbit region length.
Parameters: $\lambda=10$ meV and $T_0=1$ K.\label{fig:Ss}}
\end{figure}

Notably, the intrinsic thermopower of Eq.~\eqref{eq_s2} drops out
from the subband Seebeck coefficient in Eq.~\eqref{eq_Ss2}.
$S_s$ depends only on the transmission probabilities to cross the spin-orbit region
and is thus a purely transport property.
Therefore, we expect a stronger energy dependence of the subband thermopower
as compared with its charge analog, Eq.~\eqref{eq_s}. This is confirmed in our numerical
simulations. In Fig.~\ref{fig:Ss} we represent $S_s$ as a function of $E_F$.
The energy variation of the subband thermopower becomes more pronounced for wider
spin-orbit regions because the subband resolved transmissions differ strongly as the region
size enhances. In addition, we observe a sign change of $S_s$, implying
that for a positive difference of temperatures a positive or negative subband potential
is generated depending on the Fermi energy.
As expected, for high energies electrons are insensitive to the Rashba scattering potential and
the subband thermopower tends to zero.

\section{Conclusions}\label{sec:conclu}

We have investigated the electric and thermoelectric properties of a graphene monolayer with inhomogeneous Rashba
spin-orbit interaction patterned as a stripe along the sample. We have discussed the energy splittings due to the Rashba coupling
and their effect in the transmission probabilities. Importantly, the existence of a critical angle for only one of the two subband
states leads to a finite polarization when the externally modulated Fermi energy is of the order of the spin-orbit strength.

We have found that our system is more sensitive to temperature shifts than to potential differences.
Surprisingly enough, the thermopower is dominated by an intrinsic term which is independent of the scattering potential.
The strong energy variation is recovered when the thermopower is calculated in the subband space.
Then, an applied temperature bias creates a subband polarization, which can attain significant values
(positive or negative) at low Fermi energies.

We have considered a system free of disorder or scattering centers additional to the spin-orbit coupling.
In a realistic sample, diffusion processes should be taken into account. However, it is remarkable
that in the diffusive regime a similar intrinsic thermopower ($S\sim k_B^2 T_0/eE_F$) is obtained.\cite{zue09}
Therefore, further work is needed to clarify the behavior of the Seebeck coefficient
in the transition from the diffusive regime to the ballistic (quantum) regime considered here.
Another interesting route would focus on the role of phonons.\cite{munoz} However, we do not expect that our results
will change qualitatively since the phonon contribution is negligible at the low temperatures
considered in our work.

Our results might be tested in a suspended graphene sample with a central section deposited
onto a metallic substrate inducing a spin-orbit interaction. The coupling between the monolayer and
the metal can be tuned with an external electric field. Then, thermovoltages and thermocurrents
would be detected upon local heating of a sample side.
An alternative measurement would consider heating currents generated in response to an applied
electric current under vanishing thermal gradients (Peltier effect). Due to reciprocity,
the measured response can be related to the thermopower. Finally, hot electrons
can originate from sample irradiation, as recently demonstrated in Ref.~\onlinecite{gab11}.
Our results are thus relevant for the exciting area that emphasizes the interplay
between spin interactions and thermoelectric effects in graphene and related nanostructures.

\section*{Acknowledgments}

This work has been supported by MINECO under Grant No.~FIS2011-23526.


\begin{thebibliography}{99}
\bibitem{note}
The $1/E_F$ divergence at very low energies below $k_B T_0$ is just an artifact
of our Sommerfeld expansion, which is exact up to $(k_B T_0/E_F)^4 $. Near the charge neutrality point, $E_F$ is of the order of $K_BT_0$ and the Sommerfeld expansion breaks down.
\end{thebibliography}
\end{document}